\documentclass[twocolumn,aps,superscriptaddress,showpacs,nofootinbib,floatfix]{revtex4}

\usepackage{epsfig,bm,feynmf}

\usepackage{graphics}

\usepackage[normalem]{ulem}  % \sout{old text} for strikeout
\usepackage[dvips]{color} % For blue in-text comments and additions

\renewcommand\sout{\bgroup \color{red} \ULdepth=-.5ex \ULset}
\begin{document}

\title{Jet-dilepton conversion in expanding quark-gluon plasma}

\author{Yong-Ping Fu and Qin Xi \\Department of Physics, Lincang Teachers College, Lincang 677000, China}
%\email{ynufyp@sina.cn}

%\affiliation{Department of Physics, Lincang Teachers College, Lincang 677000, China}

%\author{Qin Xi}
%\affiliation{Department of Physics, Lincang Teachers College, Lincang 677000, China}

\date{\today}

\begin{abstract}
We calculate the production of large mass dileptons from the jet-dilepton conversion in the expanding quark-gluon plasma at Relativistic Heavy Ion Collider (RHIC) and Large Hadron Collider (LHC) energies. The jet-dilepton conversion exceeds the thermal dilepton production and Drell-Yan process in the large mass region of 3.9 GeV$<M<$5.8 GeV and 6.3 GeV$<M<$8.7 GeV in central Pb+Pb collisions at $\sqrt{s_{NN}}$=2.76 TeV and 5.5 TeV, respectively. We present the numerical solution of ideal fluid hydrodynamics. We find that the transverse flow leads to a rapid cooling of the fire ball. The suppression due to transverse flow is appeared from small to large mass, the transverse flow effect becomes important at LHC energies. The energy loss of jets in the hot and dense medium is also included.
\end{abstract}


\maketitle

\section{Introduction}

One of the most important aim in the experiments of relativistic heavy-ion collisions is that of the study of a quark-gluon plasma (QGP). The electromagnetic radiation is considered to be a useful probe for the investigation of the evolution of the QGP due to their very long mean free path in the medium \cite{shuryak}.

In relativistic heavy-ion collisions dileptons are produced from several sources. These include the dileptons from the Drell-Yan process of primary partons \cite{DY}, thermal dileptons from the interactions of thermal partons in the QGP \cite{thermal-dilepton1} and the hadron interactions in the hadronic phase \cite{HG1,HG2}, and dileptons from the hadronic decays occurring after the freeze-out \cite{HG3}. Energetic jets produced via the parton scattering in relativistic heavy-ion collisions also provide an excellent tool that enables tomographic study of the dense medium \cite{HT1,GLV1}. In Refs. \cite{jet-dilepton4} the authors indicated that the electromagnetic radiation from jets interacting with the QGP is a further source. The authors in Refs. \cite{jet-dilepton4,jet-dilepton3,AMY2} studied the production of high-energy photons from a jet passing through a QGP. The contribution of photons from the jet-photon conversion in the medium is as large as the photon yield from hard scatterings in the momentum range $p_{\bot}\approx$2$\sim$6 GeV at RHIC. The energy loss effects for jets before they convert into photons have been investigated by Turbide $\emph{et al}$, and the model of the energy loss was introduced into the calculation of the jet-dilepton conversion. The jet-dilepton conversion in the 1+1 dimensional (1+1 D) evolution of the plasma has been investigated by Refs. \cite{jet-dilepton5,AMY1,jet-dilepton2}.

In relativistic heavy-ion collisions, the relativistic hydrodynamical equations can describe the collective properties of the strongly interacting matter. The Bjorken solution provides an estimate of the 1+1 D cylindrical expansion of the plasma \cite{Bjorken}. The transverse flow effects have been calculated numerically which assumes cylindrical symmetry along the transverse direction and boost invariant along the longitudinal direction \cite{1+3D1,1+3D3,1+3D4,1+3D5}. After the initial proper time $\tau_{i}$ and initial temperature $T_{i}$ the system is regarded as thermalized. The system temperature $T$ are given as a function of proper time $\tau$ and radial distance $r$ by the numerical calculation of the flow. The transverse flow effect of the dilepton production from the QGP, with cylindrical symmetry, are shown to be important in the region of low invariant mass \cite{1+3D3}. In the present work, we study the effect of collective radial flow in jet-dilepton conversion. We find the transverse flow effect is also apparent at intermediate and high invariant mass at RHIC and LHC energies.

Jets crossing the hot and dense plasma will lose their energies. For high energy partons, the radiative energy loss is dominant over the elastic energy loss \cite{GW2}. The jet energy loss through gluon bremsstrahlung in the medium has been elaborated by several models: Gyulassy-Wang (GW) \cite{GW1,GW2}, Gyulassy-Levai-Vitev (GLV) \cite{GLV2,GLV3}, Baier-Dokshitzer-Mueller-Peigne-Schiff (BDMPS) \cite{BDMPS1,BDMPS2}, Guo-Wang (HT) \cite{HT2,HT3}, Wang-Huang-Sarcevic (WHS) \cite{WHS1,WHS2}, and Arnold-Moore-Yaffe (AMY) \cite{AMY2,AMY3,AMY4}. In Ref. \cite{jet-dilepton3,AMY2,AMY1} the authors use the AMY formalism to investigate the electromagnetic signature of jet-plasma interactions. The AMY formalism assumes that hard jets evolve in the hot medium according to the Fokker-Planck rate equations for their momentum distributions $dN^{jet}/dE$. Energy loss is described as a dependence of the parton momentum distribution on time. In this paper we use the WHS and BDMPS frameworks to calculate the energy loss of the momentum distribution of jets passing through the expanding QGP.

This paper is organized as follows. In Sec. II we discuss the ideal hydrodynamics equations. In Sec. III we calculate the jet production and jet energy loss. In Sec. IV we rigorously derive the production rate for the jet-dilepton conversion  by using the relativistic kinetic theory. The Drell-Yan process is also presented in Sec. V. Finally, the numerical discussion and summary are presented in Sec. VI and VII.

%\section{Formulation}
\section{ideal hydrodynamics}\label{hydro}

In this section we begin with the equation for conservation of energy-momentum
\begin{eqnarray}
%\nonumber\\
\partial_{\mu}T^{\mu\nu}=0,
\label{conservation}
\end{eqnarray}
the energy-momentum tensor of an ideal fluid produced in relativistic heavy-ion collisions is given by
\begin{eqnarray}
T^{\mu\nu}=(\varepsilon+P)u^{\mu}u^{\nu}-Pg^{\mu\nu},
\label{tensor}
\end{eqnarray}
where $\varepsilon$ is the energy density, $P$ is the pressure, and $u^{\mu}=\gamma(1,\emph{\textbf{v}})$ is the four-velocity of the collective flow, where $\gamma=1/(1-\emph{\textbf{v}}^{2})^{1/2}$. The $u^{\mu}$ satisfies the constraint $u^{2}=1$. We denote the space-time coordinate by $x^{\mu}=(t,\textbf{r})$ and the metric tensor by $g^{\mu\nu}=\mathrm{diag}(1,-1,-1,-1)$. We have $(g^{\mu\nu})^{2}=1+d$, $\delta_{ii}=d$, where Greek letters denote Lorentz indices, Latin letters denote three-vector indices, the notation $d$ stands for the dimensionality of the space. In the ideal fluid with cylindrical symmetry and boost invariant along the longitudinal direction, the fluid velocity vector $u^{\mu}$ can be written as \cite{1+3D1}
\begin{eqnarray}
u^{\mu}=\gamma_{r}(\tau,r)(t/\tau,v_{r}(\tau,r),z/\tau),
\label{velocity}
\end{eqnarray}
where we have used
\begin{eqnarray}
\gamma_{r}=\left(1-v_{r}^{2}(\tau,r)\right)^{-1/2},
\label{gama}
\end{eqnarray}
\begin{eqnarray}
\tau=(t^{2}-z^{2})^{1/2},
\label{tao}
\end{eqnarray}
we denote the space-time rapidity $\eta$ as
\begin{eqnarray}
\eta=\frac{1}{2}\ln\frac{t+z}{t-z},
\label{rapidity}
\end{eqnarray}
then the hydrodynamics equation Eq.(\ref{conservation}) for a transverse and longitudinal expansion can be written as
\begin{eqnarray}
 \frac{\partial\varepsilon}{\partial\tau}\!+\!\frac{\varepsilon\!+\!P}{\tau}\!+\!(\varepsilon\!+\!P)\!\left(\!\!\frac{\partial v_{r}(\tau,r)}{\partial r}+u^{\mu}\partial_{\mu}\ln \gamma_{r}(\tau,r)\!\right)=0.
\label{hydro}
\end{eqnarray}

\begin{table}
\centering\caption{Initial conditions of the hydrodynamical
expansion.}
\label{temperature}%\vspace{-1mm}
\begin{tabular}{rcccccccc}
\hline &&$T_{i}$(MeV) &$\tau_{i}$(fm/c)&$\tau_{c}$(fm/c)& \\
\hline
\hline
&Solutions with $v_{r}$ (r=0)\\
\hline
&RHIC& 370&$ 0.26$&$ 2.82$&\\
&LHC& 636&$ 0.088$&$ 4.44$&\\
&&845&$ 0.087$&$ 8.32$&\\
\hline
&Bjorken solutions ($v_{r}$=0) \\
\hline
&RHIC& 370&$ 0.26$&$ 3.22$&\\
&LHC& 636&$ 0.088$&$ 5.53$&\\
&&845&$ 0.087$&$ 12.96$&\\
\hline
\end{tabular}
\end{table}

\begin{figure}[h]
\includegraphics[width=1.1\linewidth]{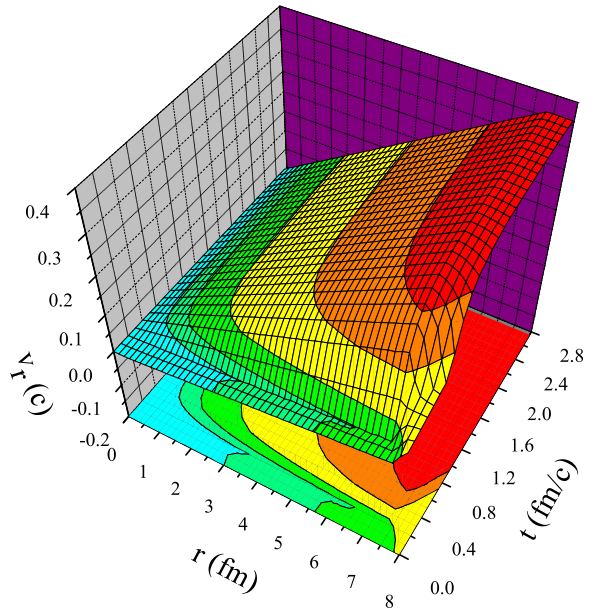}
\caption{\label{RHICvr}  (Color online)Hydrodynamical solution of the collective radial velocity for 200$A$ GeV central Au+Au collision. The black contours in the ($r,t$)-plane are the radial velocity contours, corresponding to velocity values of 0.025, 0.05, 0.1, 0.2, and 0.3 from left to right. The initial temperature $T_{i}$=370 MeV and the initial time $\tau_{i}$=0.26 fm/c, corresponding to the particle rapidity density $dN/dy$=1260. }
\end{figure}

In the case of $v_{r}=0$, Eq.(\ref{hydro}) is the well-known Bjorken equation. In a Bjorken expansion, the initial time $\tau_{i}$ and the initial temperature $T_{i}$ are related by the following
\begin{eqnarray}
T_{i}^{3}\tau_{i}=\frac{\pi^{2}}{\zeta(3)g_{Q}  }\frac{1}{\pi R^{2}_{\bot}}\frac{dN}{dy},
\end{eqnarray}
where $dN/dy$ is the particle rapidity density for the collision and $g_{Q}=42.25$ for a plasma of massless $u$, $d$, $s$ quarks and gluons. $R_{\bot}=1.2 A^{1/3}$ is the initial transverse radius of the system for a central collision. The end of the QGP phase occur at proper time $\tau_{c}=\tau_{i}\left(T_{i}/T_{c}\right)^{3}$, where $T_{c}$=160 MeV is the critical temperature of the phase transition. We use the initial temperature $T_{i}=370 $ MeV for $dN/dy=1260$ at RHIC, $T_{i}=636$ MeV for $dN/dy=2400$ at LHC (Pb+Pb, $\sqrt{s_{NN}}$=2.76 TeV), and $T_{i}=845$ MeV for $dN/dy=5624$ at LHC (Pb+Pb, $\sqrt{s_{NN}}$=5.5 TeV) \cite{jet-dilepton5,AMY1,dN/dy2,dN/dy1}. The numerical results of the initial conditions are presented in Table I. We can see that the transverse expansion leads to a more rapid cooling of the system.

\begin{figure}[h]
\includegraphics[width=1.1\linewidth]{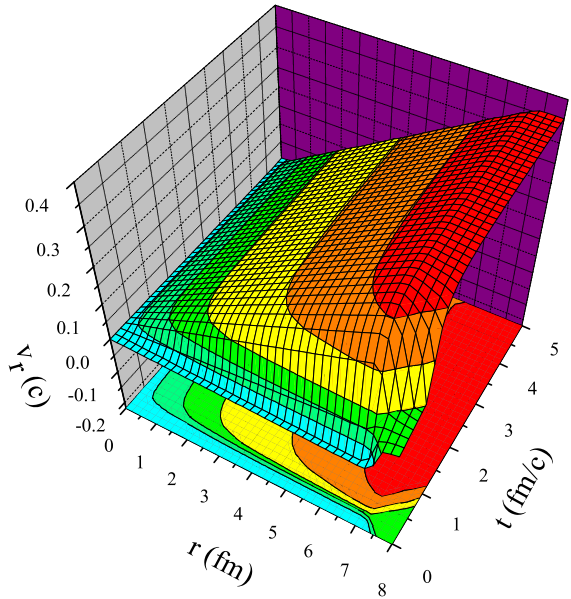}
\caption{\label{LHCvr}  (Color online)Same as Fig.\ref{RHICvr} but for cental Pb+Pb collision at $\sqrt{s_{NN}}$=5500 GeV. The initial temperature $T_{i}$=845 MeV and the initial time $\tau_{i}$=0.087 fm/c, corresponding to the particle rapidity density $dN/dy$=5624.}
\end{figure}

The initial conditions of the transverse expansion are chosen such that $v_{r}(\tau_{i},r)$=0 along with a given initial temperature $T(\tau_{i},r)=T_{i}$ within the transverse radius \cite{1+3D4}. The hydrodynamics equation (\ref{hydro}) was solved numerically using the first-order Lax finite difference scheme \cite{HG2}. We also compare with the results of earlier works \cite{hydro1,hydro2} to ensure that technical aspects are under control. To illustrate the transverse dynamics we show in Figs.\ref{RHICvr} and \ref{LHCvr} the radial velocity $v_{r}$ as a function of time and radial distance in the QGP phase. The velocity contours illustrate how the radial pressure gradient pushes the plasma to collective motion. From Figs.\ref{RHICvr} and \ref{LHCvr} we can see that as time proceeds the velocity profile $v_{r}(r)$ becomes to a nearly linear shape, and there is no acceleration near the critical time due to the disappearance of pressure gradients. These results agree well with the numerical solutions from Refs. \cite{hydro2,hydro1}.

\section{Jet energy loss}
The BDMPS model determines the energy loss of jets crossing the hot and dense plasma by means of the spectrum of energy loss per unit distance $dE/dx$. Induced gluon bremsstrahlung, rather than elastic scattering of partons, is the dominant contribution of the jet energy loss \cite{BDMPS1}. If an energetic jet paces through a long distance in the QGP, and hadronize outside the system, the energy loss of the jet is large \cite{WHS1,WHS2,GW2,BDMPS1}. However, in the case of the jet-dilepton (or jet-photon) conversion, jets travel only a short distance through the plasma before they convert into dileptons (or photons), and do not lose a significant amount of energy. The energy loss in the jet-photon conversion is found to be small, just about 20\% at RHIC \cite{jet-dilepton3}.

\begin{table}
\centering\caption{The average value of the distance covered by the jet during the passage of the jet-dilepton conversion and total distance in the expanding QGP.}
\label{temperature}%\vspace{-1mm}
\begin{tabular}{rcccccccc}
\hline &  &&$T_{i}$(MeV) &$\langle l\rangle$(fm)&$\langle \tilde{L}\rangle$(fm)& \\
\hline
\hline
& with transverse flow  \\
\hline
&RHIC&& 370 &$ 1.28$&$ >6.55$&\\
&LHC&& 636 &$ 2.18$&$  >6.66$&\\
&&& 845 &$ 4.12$&$  >6.66$&\\
\hline
&without transverse flow  \\
\hline
&RHIC&& 370 &$ 1.48$&$ 6.55$&\\
&LHC&& 636 &$ 2.72$&$ 6.66$&\\
&&& 845 &$ 6.44$&$ 6.66$&\\
\hline
\end{tabular}
\end{table}

Because we discuss the jets produced at midrapidity, in this restriction a jet will only propagate in the transverse directions.
The total distance a parton produced at $(r,\varphi)$ travels through the QGP is $\tilde{L}(r,\varphi,\tau)=\sqrt{R^{2}-r^{2}\sin ^{2 }\varphi}-r\cos \varphi$, where $R(\tau,r)$ is the radius of the expanding QGP. In the 1+1 D Bjorken evolution, neglecting the transverse expansion, the average value of $\tilde{L}=\sqrt{R_{A}^{2}-r^{2}\sin ^{2 }\varphi}-r\cos \varphi$ is $\langle \tilde{L}\rangle\approx 0.9 R_{A}$, where $R_{A}=1.2A^{1/3}$ fm is the initial radius of the system \cite{jet-dilepton5,WHS2}. Considering the transverse expansion, we have $R(\tau,r)> R_{A}$ and $\langle \tilde{L}\rangle_{(v_{r}>0)} > \langle \tilde{L}\rangle_{(v_{r}=0)} $.

In the ultra-relativistic collisions, we assume the parton is massless and travels with the speed of light in the transverse direction, as suggested in Ref. \cite{jet-dilepton5}. Then the distance of the jets passing through the QGP before the jet-dilepton conversions is
\begin{eqnarray}
l(\tau)=c(\tau-\tau_{i}),
\end{eqnarray}
where we have taken $c$=1, the average value of the distance of jet-dilepton processes is
\begin{eqnarray}
\langle l\rangle=\frac{1}{\triangle\tau}\int^{\tau_{c}}_{\tau_{i}}c(\tau-\tau_{i})d\tau=\frac{1}{2}c\triangle\tau,
\end{eqnarray}
where $\triangle\tau=\tau_{c}-\tau_{i}$ is the lifetime of the QGP phase. In Table II we can see that the distance $\langle l\rangle$ of the jet-dilepton conversion process is smaller than the total distance $\langle \tilde{L}\rangle$. The jets covers a short distance in the QGP before they convert into dileptons. Since the transverse expansion will reduce the lifetime of the QGP, the distance $\langle l\rangle_{(v_{r}>0)}$ is smaller than the value of $\langle l\rangle_{(v_{r}=0)}$ at RHIC and LHC energies.

\begin{figure}[h]
\includegraphics[width=1.1\linewidth]{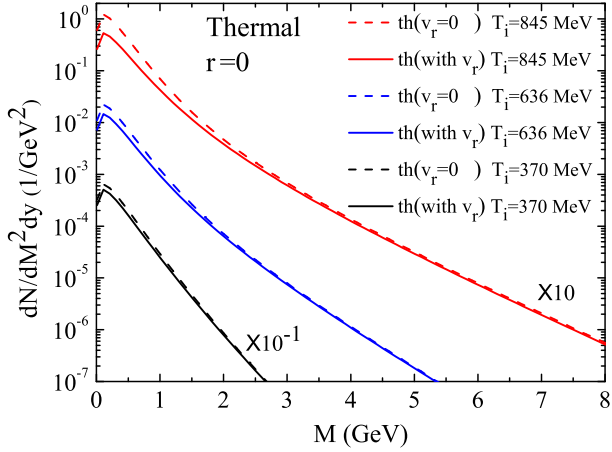}
\caption{\label{1}  (Color online)The results of thermal dileptons produced from the QGP phase at RHIC and LHC energies. In central Au+Au collisions at $\sqrt{s_{NN}}$=200 GeV the initial temperature is $T_{i}$=370 MeV. In central Pb+Pb collisions at $\sqrt{s_{NN}}$=2.76 TeV and 5.5 TeV, the initial temperature is $T_{i}$=636 MeV and $T_{i}$=845 MeV, respectively. The dashed line means the thermal dileptoons produced from the QGP without transverse flow. The solid line means thermal dileptons produced from the expanding QGP with transverse flow.}
\end{figure}

In WHS approach the authors of Ref. \cite{WHS1,WHS2} use a phenomenological model to describe the modification of the jet fragmentation function due to the energy loss. This approach is useful for studies of the parton energy loss and multiple final-state scatterings \cite{WHS2}. Given the inelastic scattering mean free path, $\lambda_{a}$, the probability for a jet to scatter $n$ times within a distance $ L$ in the hot medium can be written as \cite{WHS1}
\begin{eqnarray}
P_{a}(n)=\frac{( L/\lambda_{a})^{n}}{n!}e^{- L/\lambda_{a}}.
\end{eqnarray}
The yield $dN_{jet}/d^{2}p_{\bot}dy_{jet}$ for producing jets with energy loss in the hot medium can be written as
\begin{eqnarray}\label{jet}
\frac{dN_{jet}}{d^{2}p_{\bot}dy_{jet}}=\frac{\sum_{n=0}^{N}P_{a}(n)\left(1-\frac{n\varepsilon_{a}}{E_{\bot}}\right)\frac{dN_{jet}^{0}}{d^{2}p'_{\bot}dy_{jet}}(p'_{\bot}, L)}{\sum_{n=0}^{N}P_{a}(n)},
\end{eqnarray}
the number of inelastic scattering is limited to $N=E_{\bot}/\varepsilon_{a}$, $E_{\bot}$ is the transverse energy of the produced jet, $\varepsilon_{a}$ is the average energy loss per scattering. $p'_{\bot}$ is the transverse momentum of the parton, we have $p_{\bot}=p'_{\bot}-\triangle E$, the energy loss $\triangle E=n\varepsilon_{a}$. The energy loss per unit distance is thus $dE_{a}/dx=\varepsilon_{a}/\lambda_{a}$.

The energy-loss per unit distance in the medium of a finite size $L$ is given by BDMPS \cite{BDMPS1}:
\begin{eqnarray}
\frac{dE_{a}}{dx}=\frac{\alpha_{s}c_{a}\mu^{2}}{8\lambda_{g}}L\ln\frac{L}{\lambda_{g}},
\end{eqnarray}
where $c_{a}$=4/3 for quarks and 3 for gluon, $\mu^{2}=4\pi\alpha_{s}T^{2}(\tau,r)$, $\mu$ is the Debye mass of the medium, $\lambda_{g}=\pi\mu^{2}/\left(126\alpha_{s}^{2}\zeta(3)T^{3}(\tau,r)\right)$ is the gluon mean free path \cite{GW2}. When a very energetic parton is propagating through a hot medium and scattering $n$ times, the propagating distance is $L=n\lambda_{q}$ \cite{BDMPS1,GW2,WHS1,WHS2}, the quark mean free path is $\lambda_{q}=9\lambda_{g}/4$. For large values of $N$, the energy loss is $\triangle E
=\int_{0}^{L}\frac{dE_{a}}{dx} dx$.

\begin{figure}[h]
\includegraphics[width=1.1\linewidth]{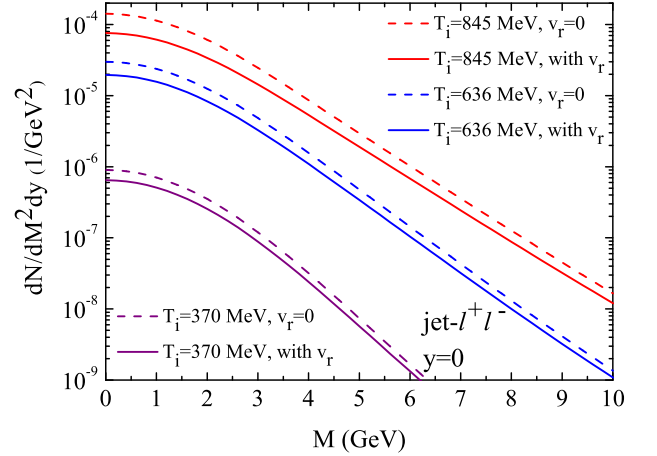}
\caption{\label{2}  (Color online)Dileptons originating from the passage of the jets passing through the QGP at RHIC and LHC energies. Dashed line: jet-dilepton conversion without the transverse flow; solid line: jet-dilepton conversion with the transverse flow. The jet energy loss is included.}
\end{figure}

The initial yield $dN^{0}_{jet}/d^{2}p'_{\bot}dy_{jet}$ for producing jets in the relativistic heavy-ion collisions ($A+B\rightarrow jets+X$) can be factored in the perturbative QCD
(pQCD) theory as \cite{Owen}
\begin{eqnarray}\label{eq8}
\frac{dN^{0}_{jet}}{d^{2}p'_{\bot}dy_{jet}}&\!\!\!\!=\!\!\!&T_{AA}\sum_{a,b}\!\frac{1}{\pi}\!
\!\!\int_{x_{a}^{min}}^{1}\!\!dx_{a}G_{a/A}(x_{a},Q^{2})G_{b/B}(x_{b},Q^{2}) \nonumber\\[1mm]
&& \times\frac{x_{a}x_{b}}{x_{a}-x_{1}}K_{jet}\frac{d\hat{\sigma}_{ab\rightarrow
cd}}{d\hat{t}},
\end{eqnarray}
where $T_{AA}=9A^{2}/8\pi R^{2}_{A}$ is the nuclear thickness
for central collisions \cite{jet-dilepton3,jet-dilepton4}. $x_{a}$ and $x_{b}$ are the momentum fraction of the parton. The momentum fractions with the rapidity
are given by $
x_{a}^{min}=x_{1}/(1-x_{2})$ and $x_{b}=x_{a}x_{2}/(x_{a}-x_{1})$,
where the variables are $x_{1}=x_{T}e^{y_{jet}}/2$, $x_{2}=x_{T}e^{-y_{jet}}/2$,
$x_{T}=2p'_{\bot}/\sqrt{s_{NN}}$. $\sqrt{s_{NN}}$ is the center of mass
energy of the colliding nucleons.
The parton distribution for the
nucleus is given by
\begin{eqnarray}\label{eq11}
G_{a/A}(x_{a},Q^{2})&=&R^{a}_{A}(x_{a},Q^{2})[Zf_{a/p}(x_{a},Q^{2})  \nonumber\\[1mm]
&&  +(A-Z)f_{a/n}(x_{a},Q^{2})]/A,
\end{eqnarray}
where $R^{a}_{A}(x_{a},Q^{2})$ is the nuclear modification of the
structure function \cite{shadowing}, $Z$ is the number of protons, $A$
is the number of nucleons. The functions $f_{a/p}(x_{a},Q^{2})$ and
$f_{a/n}(x_{a},Q^{2})$ are the parton distributions of the proton
and neutron, respectively \cite{parton}.
$d\hat{\sigma}_{ab\rightarrow cd}/d\hat{t}$ is the cross section of
parton collisions at leading order, these processes are:
$q\bar{q}\rightarrow q'\bar{q}'$, $qq'\rightarrow qq'$,
$q\bar{q}'\rightarrow q\bar{q}'$, $qq\rightarrow qq$,
$q\bar{q}\rightarrow q\bar{q}$, $qg\rightarrow qg$,
and $gg\rightarrow q\bar{q}$ \cite{parton2}. $K_{jet}$ is the pQCD
correction factor to take into account the next-to-leading order
(NLO) effects, we use $K_{jet}=$ 1.7 for RHIC and 1.6 for LHC
\cite{AMY2}.

\begin{figure}[h]
\includegraphics[width=1.1\linewidth]{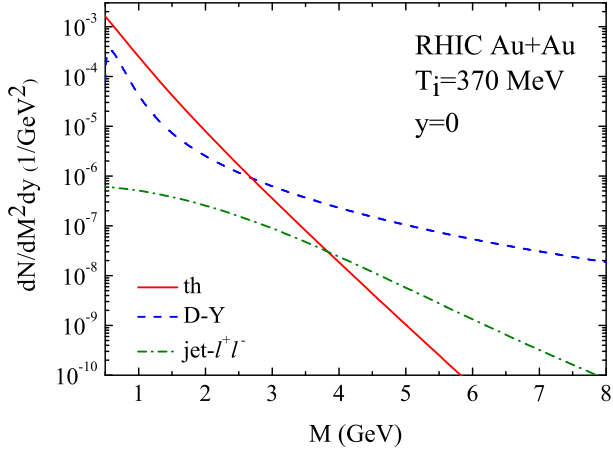}
\caption{\label{4}  (Color online)Dilepton yield for central Au+Au collisions at $\sqrt{s_{NN}}$=200 GeV. Solid line: thermal dileptons produced from the expanding QGP; dashed line: Drell-Yan contribution; dash-dot line: dileptons produced by the jet-dilepton conversion. The energy loss and transverse flow effects are considered. }
\end{figure}

\section{Jet-dilepton conversion}

The jets passing through the QGP can produce large
mass dileptons by annihilation with the thermal partons of the hot medium
($q_{jet}\bar{q}_{th}\rightarrow l^{+}l^{-}$ and
$q_{th}\bar{q}_{jet}\rightarrow l^{+}l^{-}$). By using the
relativistic kinetic theory, the production rate for the above
annihilation process can be written as \cite{jet-dilepton2}
\begin{eqnarray}\label{eq1}
R_{jet-l^{+}l^{-}}\!\!\!=\!\!\!\int\frac{d^{3}p_{1}}{(2\pi)^{3}}\int\frac{d^{3}p_{2}}{(2\pi)^{3}}
f_{jet}(\emph{\textbf{p}}_{1})f_{th}(\emph{\textbf{p}}_{2})\sigma(M)v_{12}.
\end{eqnarray}
The cross section of the $q\bar{q}\rightarrow l^{+}l^{-}$
interaction is given by $\sigma(M)=4\pi\alpha^{2}N_{c}N_{s}^{2}\sum_{q}e_{q}^{2}/3 M^{2}$, where the parameters $N_{c}$ and $N_{s}$ are the color number and
spin number, respectively. The relative velocity is $v_{12}=(p_{1}+p_{2})^{2}/2E_{1}E_{2}$. In the relativistic collisions, $|\emph{\textbf{p}}|\approx E$, the
integration over
$d^{3}p=|\emph{\textbf{p}}|^{2}d|\emph{\textbf{p}}|d\Omega$ can be
done with the relatively simple result \cite{jet-dilepton2}
\begin{eqnarray}\label{eq5}
\frac{dR_{jet-l^{+}l^{-}}}{dM^{2}}=\frac{\sigma(M)M^{2}}{2(2\pi)^{4}}
\int dp_{\bot}
f_{jet}(p_{\bot})Te^{-\frac{M^{2}}{4p_{\bot}T}},
\end{eqnarray}
The jets produced in initial parton collisions are defined by all
partons with transverse momentum $p_{\bot}^{jet}\gg$1 GeV
\cite{jet-dilepton5}. The dilepton production is sensitive to the
choice of the cutoff $p_{\bot}^{jet}$. In order to avoid such
sensitivity, the authors of Ref. \cite{jet-dilepton5,AMY1} have constrained a
lower cutoff $p_{\bot}^{jet}\geq$4 GeV. We
adopt this limit in the integration of Eq.(\ref{eq5}).

\begin{figure}[h]
\includegraphics[width=1.1\linewidth]{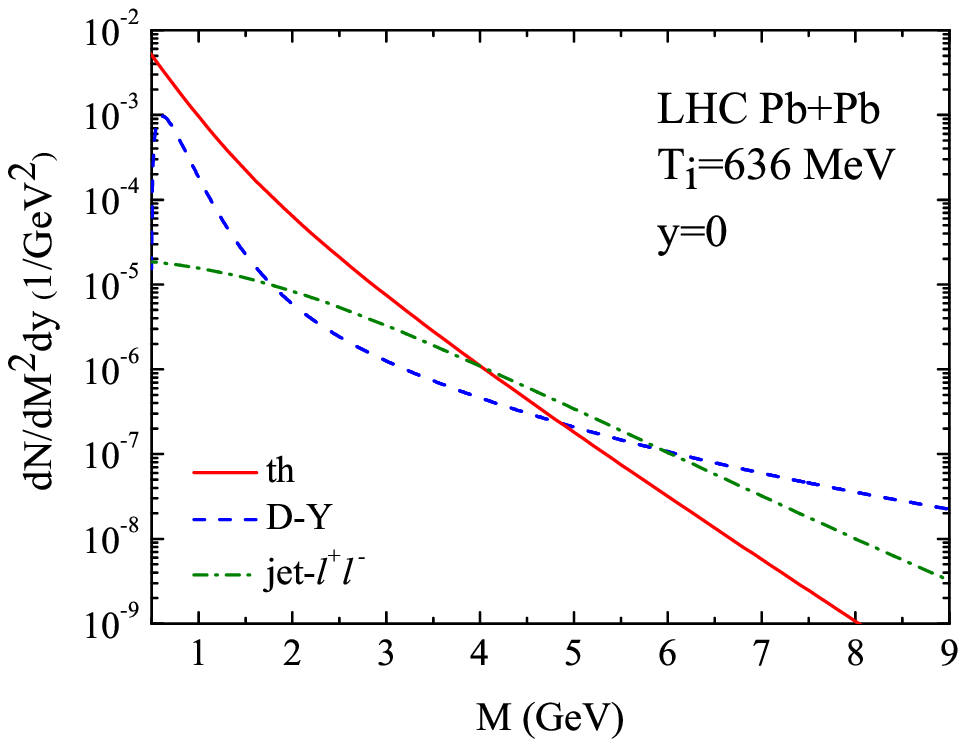}
\caption{\label{6}  (Color online)Same as Fig.\ref{4} but for central Pb+Pb collisions at $\sqrt{s_{NN}}$=2.76 TeV.}
\end{figure}

The phase-space distribution function for a jet, assuming the constant transverse density of nucleus, is as follows \cite{jet-dilepton1}
\begin{eqnarray}\label{fjet}
f_{jet}=\frac{(2\pi)^{3}}{g_{q}\pi R_{\bot}^{2}\tau p_{\bot}}\frac{dN_{jet}}{d^{2}p_{\bot}dy_{jet}}(y_{jet}=0),
\end{eqnarray}
where $g_{q}=6$ is the spin and color degeneracy of the quarks (and antiquarks).

If the phase-space distribution for the quark jets
$f_{jet}(\emph{\textbf{p}})$ is replaced by the thermal distribution
$f_{th}(\emph{\textbf{p}})$ in Eq.(\ref{eq5}), one can obtain the
yield for producing thermal dileptons as \cite{thermal-dilepton1}
\begin{eqnarray}\label{eq6}
\frac{dN_{th}}{dM^{2}dy}\!=\!\frac{4\alpha ^{2} M \sum e _{q} ^{2}}{(2\pi)^{3}}\!\!\int\!\!\tau d\tau\!\!\int\! \!d^{2}r
T(\tau,r)K_{1}\!\!\left(\!\!\frac{M}{T(\tau,r)}\!\!\right),
\end{eqnarray}
where the Bessel function is $K_{1}(z)=\sqrt{\pi/(2z)}e^{-z}$.

Because we are interested in jets produced at midrapidity ($y_{jet}$=0), we only consider dileptons produced at midrapidity ($y$=0). The dileptons produced from the passage of jet passing through the QGP are finally obtained from Eqs.(\ref{jet}), (\ref{eq5}) and (\ref{fjet}) with the space-time integration and transverse momentum integration \cite{jet-dilepton2,AMY1}. After some algebra, we get
\begin{eqnarray}
\frac{dN_{jet-l^{+}l^{-}}}{dM^{2}dy}\!\!&=&\!\!\frac{\sigma(M)M^{2}}{2(2\pi)^{4}}\!\!\int\!\!\tau d\tau\!\!\int\! \!d^{2}r\!\!\int \!\! dp_{\bot}T(\tau,r)e^{-\frac{M^{2}}{4p_{\bot}T(\tau,r)}}\nonumber\\[1mm]
&& \times \frac{(2\pi)^{3}}{g_{q}\pi R_{\bot}^{2}\tau p_{\bot}}\frac{dN_{jet}}{d^{2}p_{\bot}dy_{jet}}|_{y_{jet}=0}.
\end{eqnarray}
In the jet-dilepton conversion processes the jet only propagates in the pure QGP phase, therefore we limit the $\tau$ integration as $[\tau_{i},\tau_{c}]$.

\begin{figure}[h]
\includegraphics[width=1.1\linewidth]{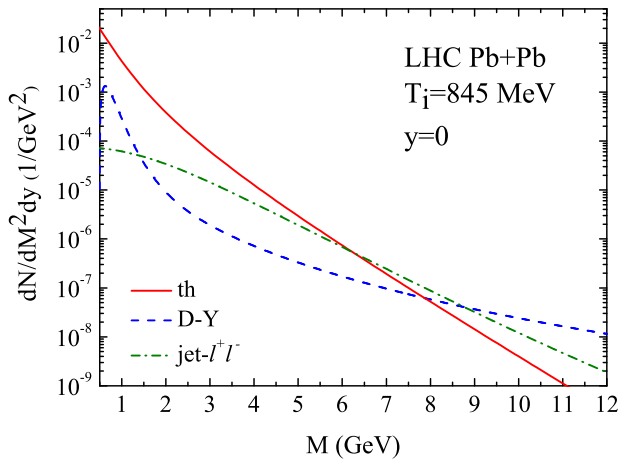}
\caption{\label{8}  (Color online)Same as Fig.\ref{4} but for central Pb+Pb collisions at $\sqrt{s_{NN}}$=5.5 TeV.}
\end{figure}

\section{Drell-Yan process}
In the central collisions of two equal-mass nuclei with mass number
$A$ the yield for producing Drell-Yan pairs with the invariant mass
$M$ and rapidity $y$ can be obtained as \cite{DY},
\begin{eqnarray}\label{eq13}
\frac{dN_{DY}}{dM^{2}dy}&\!\!\!\!\!\!=\!\!\!\!\!\!&T_{AA}K_{DY}\frac{4\pi\alpha^{2}}{9M^{4}}\sum_{q}e_{q}^{2}
[x_{a}G_{q/A}(x_{a},Q^{2})   \nonumber\\[1mm]
&&\times x_{b}G_{\bar{q}/B}(x_{b},Q^{2})+(q\leftrightarrow\bar{q})],
\end{eqnarray}
where the momentum fractions with rapidity $y$ are $x_{a}=M e^{y}/\sqrt{s_{NN}}$, $x_{b}=M e^{-y}/\sqrt{s_{NN}}$.
A $K_{DY}$ factor of 1.5 is used
to account for the NLO corrections \cite{jet-dilepton2,thermal-dilepton2}.

\section{Results and Discussions}\label{results}

In Fig.\ref{1} we plot the results of thermal dileptons produced from the QGP at RHIC and LHC energies. In the central Au+Au collisions at  $\sqrt{s_{NN}}$=200 GeV we choose the initial temperature of the expanding QGP $T_{i}$=370 MeV \cite{jet-dilepton5,AMY1}. Then we have the initial time $\tau_{i}$=0.26 fm/c and the critical time $\tau_{c}$=2.82 fm/c corresponding to $y$=0 and $r$=0. In the 1+1 D Bjorken expansion the critical time \cite{1+3D5} $\tau_{c}$ is 3.22 fm/c at RHIC. The life time of the QGP phase with the transverse expansion ($\triangle\tau_{(v_{r}>0)}$=2.56 fm/c) is smaller than the one of the Bjorken case ($\triangle\tau_{(v_{r}=0)}$=2.96 fm/c) at RHIC energy. At LHC we have $\triangle\tau_{(v_{r}>0)}$=4.352 fm/c, $\triangle\tau_{(v_{r}=0)}$=5.442 fm/c and $\triangle\tau_{(v_{r}>0)}$=8.233 fm/c, $\triangle\tau_{(v_{r}=0)}$=12.873 fm/c corresponding to $T_{i}$=636 MeV and $T_{i}$=845 MeV, respectively. The initial conditions at RHIC and LHC are calculated in Table I. We observe that transverse flow effect of the expansion leads to a rapid cooling of the fire ball, especially at LHC energies.

\begin{figure}[h]
\includegraphics[width=1.1\linewidth]{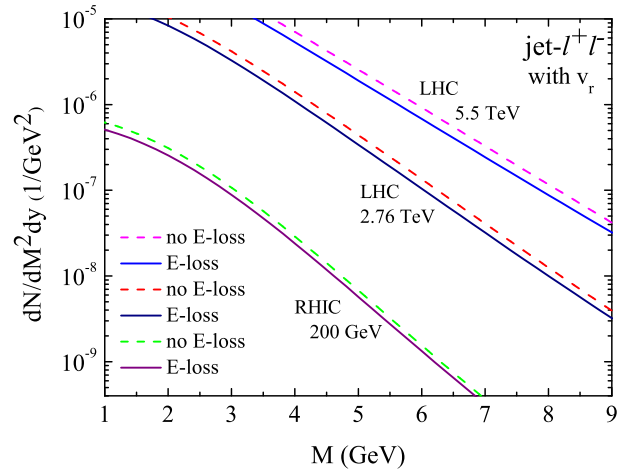}
\caption{\label{9}  (Color online)Effect of jet energy loss on the jet-dilepton conversion at RHIC and LHC energies. The solid lines include the jet energy loss, while the dashed lines do not.}
\end{figure}

For comparison, the yields of the thermal dileptons and the jet-dilepton conversion from the 1+1 D Bjorken expansion and cylindrical expansion with transverse flow are given in Figs.\ref{1} and \ref{2}, respectively. In Fig.\ref{1} we find that the transverse flow effect reduces the yields from low to high invariant mass and the reduction is largest at small $M$, the transverse flow effect is still apparent at intermediate $M$. The reduction of thermal dileptons is in the region of $M<$2.5 GeV for $T_{i}$=370 MeV, $M<$4 GeV for $T_{i}$=636 MeV and $M<$6 GeV for $T_{i}$=845 MeV. The thermal production is suppressed by a factor$\sim$ 2 at $M\sim$ 0.5 GeV for $T_{i}$=845 MeV. In Fig.\ref{2} the reduction of the jet-dilepton conversion due to the transverse flow is apparent from low to high invariant mass. The reduction of the jet propagating length due to the transverse flow leads to the yield suppression at large mass. We find a factor$\sim$ 1.5 of suppression at $M\sim$ 6 GeV for $T_{i}$=845 MeV. The transverse flow effect becomes important at the QGP phase at the LHC energies.

Figs.\ref{4}, \ref{6} and \ref{8} present the results for thermal dileptons, direct dileptons from Drell-Yan process and dileptons from the interaction of jets with the expanding QGP at RHIC and LHC energies, respectively. In Fig. \ref{4} the contribution of the jet-dilepton conversion is not prominent at RHIC energy. However the jet-dilepton conversion is comparable to that of the thermal contribution and Drell-Yan process at LHC energies.
The jet-dilepton conversion is a dominant source in the region of 3.9 GeV$<M<$5.8 GeV and 6.3 GeV$<M<$8.7 GeV in central Pb+Pb collisions at $\sqrt{s_{NN}}$=2.76 TeV and 5.5 TeV, respectively (see Figs. \ref{6} and \ref{8}). In Ref. \cite{jet-dilepton2} the contribution of the jet-dilepton conversion is prominent in the region for 4 GeV$<M<$10 GeV at LHC ($\sqrt{s_{NN}}$=5.5 TeV) due to the absence of the transverse flow and the energy loss effects.

The energy loss effect on jet-dilepton conversion is presented in Fig. \ref{9} at RHIC and LHC energies. The energy loss effect suppresses the jet-dilepton spectrum, the suppression decreases with increasing invariant mass $M$. For a given invariant mass $M$ and thermal parton energy $E_{th}$, the minimum energy of the jet is $E_{jet}=M^{2}/4E_{th}$ \cite{jet-dilepton2}. The energy loss rate $\triangle E/E_{jet}\propto M^{-2}$, this implies that jet-dilepton conversion with large $M$ favors small jet propagation length and small energy loss. The energy loss depends on the propagating length $L$ of the jet. In Table II we find that $\langle l\rangle_{\textrm{LHC}}>\langle l\rangle_{\textrm{RHIC}}$, the large value of the distance corresponds to the increase of the energy loss rate. The suppression induced by the jet energy loss is much larger at LHC energies. At RHIC dileptons are reduced by about 20$ \% $ for $M$=1 GeV, and 16$ \% $ for $M$=4 GeV. These results agree with the numerical results from the AMY approach \cite{AMY1,AMY2}. In central Pb+Pb collisions at $\sqrt{s_{NN}}$=2.76 TeV the suppression is about 21$ \% $ and 20$ \% $ at $M$=4 GeV and 6 GeV, respectively. At LHC($\sqrt{s_{NN}}$=5.5 TeV) the suppression is about 26$ \% $ at $M$=6 GeV and 23$ \% $ at $M$=9 GeV.

The main background for the dilepton production in the intermediate and high invariant mass region is the decay of open charm and bottom mesons.
The $c\bar{c}$($b\bar{b}$) pairs are produced from the initial hard
scattering of partons and can thereafter fragment into $D$($B$) and
$\bar{D}$($\bar{B}$) mesons. If the energy loss of heavy quarks
crossing the hot medium is considered, the contribution of the decay
of open charm and bottom mesons will be suppressed \cite{CEL1,CEL2}. In Ref.\cite{CEL3} the authors study the Parton-Hadron-String Dynamics transport approach, and find that the contribution of the dileptos from the the decays of $D\bar{D}$ and $B\bar{B}$ mesons is lower than the thermal dileptons in the intermediate and high invariant mass region at LHC. This provides the possibility to measure the jet-dilepton conversion from the QGP. Since there is no single model that could address reliably decays of open charm and bottom mesons,
these backgrounds are not plotted in this article and the background
of $J/\Psi$ vector meson decay is also not concerned.

\section{Summary}
We have calculated the large mass dilepton produced from the jet-dilepton conversion, QGP and Drell-Yan at RHIC and LHC energies. We presented the numerical solutions of the ideal hydrodynamics equation with cylindrical symmetry and boost invariant along the longitudinal direction. We have found that the transverse flow effect from the expanding QGP leads to a smaller life time of the QGP phase, and suppresses the production of jet-dileptons and thermal dileptons from low to high invariant mass region. We have found an important window of the jet-dilepton conversion. The jet-dilepton conversion is a dominant source in high invariant mass regions at LHC, after the background of heavy quark decays is subtracted. The jet energy loss has been included by using the WHS and BDMPS frameworks. The jet energy loss is relatively small at RHIC due to the small propagating length of a jet.

\section*{Acknowledgements}

We thank D. K. Srivastava for original suggestions. This work was supported by Science Foundation of LTC under Project No. LCSZL2013004, and Scientific Research Foundation of the Education Department of Yunnan Province of China under Project No. 2012Y274.

%%%%%%%%%%%%%%%%%%%%%%%%% References

\end{document}